\documentclass{article}

\usepackage{arxiv}

\usepackage[utf8]{inputenc} 
\usepackage[T1]{fontenc}    
\usepackage{url}            
\usepackage{booktabs}       
\usepackage{nicefrac}       
\usepackage{microtype}      
\usepackage{lipsum}		
\usepackage{graphicx}
\usepackage{natbib}
\usepackage{doi}
\usepackage{amsmath,amsfonts,amssymb}
\usepackage{setspace}
\usepackage{tocloft}
\usepackage{float}
\usepackage[table]{xcolor}
\usepackage{macro_ref}
\usepackage{caption}
\usepackage{subcaption}
\usepackage{lineno}
\usepackage{soul}

\title{Achieving efficient broadband spatial filtering for LIFE: status and plan}

\author{}

\date{}

\renewcommand{\shorttitle}{Garreau et al.: Achieving efficient broadband spatial filtering for LIFE: status and plan}

\hypersetup{
pdftitle={Achieving efficient broadband spatial filtering for LIFE: status and plan},
pdfsubject={astro-ph.IM},
pdfauthor={Germain Garreau},
pdfkeywords={nulling interferometry; infrared instrumentation; exoplanets; LIFE},
}

\begin{document}
\maketitle
\vspace{-6em}
\begin{center}
G. Garreau$^{a}$\footnote{ggarreau@phys.ethz.ch},
T. Birbacher$^{a}$,
L. Desdoigts$^{b,c}$,
L. D. Feinberg$^{a}$,
A. M. Glauser$^{a}$,
J. T. Hansen$^{a}$,
M. Ireland$^{d}$,
J. Pino$^{a}$,
E. Spalding$^{a}$,
A. K. Taras$^{b}$, and
S. P. Quanz$^{a,e}$
~\newline

$^a$ETH Zürich, Institute for Particle Physics \& Astrophysics, Wolfgang-Pauli-Str. 27, 8093 Zürich, Switzerland\\
$^b$Leiden Observatory, Leiden University, PO Box 9513, 2300 RA Leiden, The Netherlands\\
$^c$Sydney Institute for Astronomy, School of Physics, University of Sydney, Camperdown, NSW 2006, Australia\\
$^d$Research School of Astronomy \& Astrophysics, Australian National University, Canberra, ACT 2611, Australia\\
$^e$ETH Zürich, Department of Earth and Planetary Sciences, Sonneggstrasse 5, 8092 Zürich, Switzerland
\end{center}
~\vspace{-0.5em}

\begin{abstract}
    Nulling interferometry is one of the most promising techniques that is envisioned for the imaging and characterization of exoplanets in the mid-infrared for ground-based and space-based observatories. 
    On the ground, the upcoming Asgard/NOTT visitor instrument for the Very Large Telescope Interferometer (VLTI) is expected to be the first nuller to observe young giant exoplanets.
    The Large Interferometer For Exoplanets (LIFE) project aims at implementing long-baseline nulling interferometry in space to image and characterize Earth-like exoplanets. 
    LIFE requires to reach deep ($<10^{-5}$) null depths over a large bandwidth in the mid-infrared (MIR: 4-18.5$\,\mu$m) with a high throughput ($>15\,\%$).
    These requirements are necessary to detect and characterize the thermal emission of Earth-like exoplanets. To achieve deep null depths, a spatial filter is necessary to wash away the wavefront aberrations that would otherwise be a limiting factor for the contrast. However, efficient spatial filtering with high throughput ($>$95\,\%) is challenging to achieve over such a large bandwidth.
    In this study, we explore the possibility of broadband spatial filtering using two step-index fibers previously studied for the Darwin mission proposal: Te-As-Se chalcogenide (TAS) and silver halide (AgBr) fibers. Using $\partial$Lux, we also simulate the performance of phase-induced amplitude apodization (PIAA) with aspherical mirrors to achromatically apodize the pupil plane of a beam and improve its coupling efficiency in both fibers. 
    The results show that a broadband geometric coupling efficiency of $>95\,\%$ can be achieved, with a manufacturing precision of $<100\,$nm for the PIAA mirrors. An achromatic apodization of the beams for LIFE is therefore compatible with a number of spectral channels of $\geq2$, defined by the number of spatial filters used.
\end{abstract}


\section{INTRODUCTION}\label{sec:introduction}
The characterization of exoplanets atmosphere using a space-based mid-infrared nulling interferometer has was proposed first in 1978 \citep{bracewell_detecting_1978}, and later in the 1990s \citep{AngelWolf1997}. In the early 2000s, twin projects for such space missions were supported by both NASA \citep[TPF-I: ][]{Beichman1999, Lawson2007} and ESA \citep[\textit{Darwin}: ][]{Leger1996, Kaltenegger2005, Cockell2009}. 
The Large Interferometer For Exoplanets \citep[LIFE: ][]{Quanz2022} is a space-mission proposal that builds on the heritage of TPF-I/\textit{Darwin}, proposing to perform nulling interferometry with a formation-flying constellation of four collector spacecrafts and one combiner spacecraft, with a spectral range of 4-18.5$\,\mu$m.
This waveband of more than two octaves in the mid-infrared is necessary to observe the absorption bands of more molecules (e.g., CO$_2$, NH$_3$, CH$_4$, ...) which would potentially populate the atmosphere of Earth-like planets, and the spectral fingerprint of potential bio-signatures \citep{Konrad2022}. Updates on the progress of LIFE are presented in these proceedings (Glauser et al. 14148-64).

The technical feasibility of LIFE is currently investigated by the Nulling Interferometry Cryogenic Experiment \citep[NICE: ][]{Ranganathan2024,Birbacher2026}, a testbed hosted at ETH Zürich. NICE is currently in its warm phase, demonstrating $<10^{-5}$ null depths at 4.7\,$\mu$m with 22\,\% throughput \citep{Birbacher2026} (see Hansen et al. 14148-26 in these proceedings). In these proceedings, an update is also given on the cryogenic characterization of elements for NICE (Hansen et al. 14154-97) and on the strategy for its control and compensation mechanisms (Birbacher et al. 14148-119). 

One of the key aspects of LIFE is the choice of spatial filters to clean the phase aberrations of the wavefront. Spatial filters are often used in interferometry to relax the requirements on wavefront quality; the first implementation of such filters using single-mode fibers was for the IOTA/FLUOR instrument \citep{coude_du_foresto_fluor_1998}. They become critical to achieve null depths of $<10^{-5}$ for the shorter wavelengths of LIFE. However, the losses due to injection and propagation of light in the spatial filter can quickly dominate the throughput loss of the instrument.

Phase-induced amplitude apodization \citep[PIAA: ][]{Guyon2003, Guyon2005} has been proposed for exoplanets imaging to achieve high-contrast ($10^9$) at $<2\lambda/D$ angles, with $\lambda$ the wavelength, and $D$ the diameter of the primary mirror. PIAA has first been implemented with aspherical lenses at Subaru/SCExAO \citep{Lozi2009,Jovanovic2015,Jovanovic2017} to improve the injection of the light into single-mode fibers.
A previous study \citep{Ireland2024} proposed to combine PIAA and the use of a single endlessly single-mode photonic crystal waveguide \citep{Knight1996,Birks1997} to achieve efficient spatial filtering across 8-18\,$\mu$m for LIFE. In this study, an alternative strategy is proposed to achieve high geometric coupling efficiency ($>95\,\%$) across the full LIFE bandwidth combining PIAA and multiple step-index single-mode fibers. Section\,\ref{sec:method} describes the method of the study, including the fibers and the PIAA system considered. Section\,\ref{sec:results} presents the results of the simulation, the broadband coupling efficiencies obtained, and the derived tolerancing of the PIAA optics.

\section{Method}\label{sec:method}
The method here first consists of choosing the spatial filters we want to use in our model, and their chromatic behavior. Then we need to choose how to implement the PIAA to simulate the beam that will be filtered, and to get its coupling efficiency with the spatial filter.

\subsection{Step-index fiber options}\label{sec:fibers}
In the context of Darwin/TPF-I, different options of spatial filters have been investigated. Two options stand out as the most mature step-index single-mode fibers in the mid-infrared:

\paragraph{Te-As-Se chalcogenide (TAS)} TAS fibers have been investigated simultaneously for Darwin \citep{Faber2006,Houizot2007,Cheng2009} and TPF-I \citep{Aggarwal2002,Lawson2017}. They provide relatively low propagation losses ($<0.1$\,dB/cm) in the 4 to 9\,$\mu$m range, with high-order cladding mode rejection that should be able to achieve the requirement of $>10^{5}$ with a fiber length of 20\,cm and with an appropriate coating of the cladding (B. Bureau, private communication). 

\paragraph{Silver halide (AgBr)} AgBr fibers have been mainly investigated in the context of Darwin \citep{Wallner2004, Ksendzov2008,Lewi2009,Flatscher2017}. AgBr has the advantage of having low propagation losses at long wavelengths ($\leq1$\,dB/m up to 18$\,\mu$m) \citep{Lewi2008}, and a high-order mode rejection of 15000 has been demonstrated at 10.6\,$\mu$m at cryo-temperatures \citep{Ksendzov2008,Flatscher2017}.

Although these fibers currently seem like the most mature technologies to perform spatial filtering with step-index fibers, additional work would still be required to meet all the requirements for LIFE. The broadband rejection ratio of TAS fibers at cryo-temperatures should be tested. Actual propagation losses of single-mode AgBr fibers in the longer wavelengths should also be measured, and improvement on the rejection ratio are also necessary. It should also be noted that, to our knowledge, the manufacturing of single-mode AgBr fibers is not available by any company/research institute at the moment.
The work presented in this study is therefore subject to uncertainty from the development of these fibers, and is likely to evolve in the future.

\subsection{PIAA simulation}\label{sec:PIAA}
The PIAA system is simulated using a package of $\partial$Lux \citep{Desdoigts2023} implemented for PIAA-Zernike Wavefront Sensor \citep{taras2026}, and also presented in these proceedings (Taras et al. 14150-186). $\partial$Lux performs physical optics simulation of astronomical imaging, and leverages the machine learning framework of JAX \citep{jax2018github} to achieve GPU acceleration and automatic differentiation. The extra package used here implements tools to create PIAA optics using either refractive or reflective surfaces, fitting any given sag surface with a mild aspheric shape \citep{Forbes2007}. The generated PIAA object can then be used to propagate light and perform the requested apodization. 

In this study, we consider a PIAA system made of two mirrors with aspherical shapes. The theoretical shapes of these mirrors -- named $M_1$ and $M_2$ -- can be found using the differential equation in \cite{Guyon2003}

\begin{equation}\label{eq:guyon2003}
    \frac{dM_1(r_1(t))}{dr_1(t)} = \frac{dM_2(r_2(t))}{dr_2(t)} = \frac{\sqrt{A(t)^2 + B(t)^2} - B(t)}{A(t)},
\end{equation}
where
\begin{align}
    A(t) &= r_2(t) - r_1(t), \\
    B(t) &= M_2(r_2(t)) - M_1(r_1(t)),
\end{align}
with $r_1(t)$ and $r_2(t)$ being the radii of the entrance and exit pupils, respectively, for a fraction of the total beam intensity $t$. 

The two free parameters to generate the PIAA are the diameters of the entrance and exit pupils $D_{en}$ and $D_{ex}$, respectively. They will determine $r_1(t)$ and $r_2(t)$ assuming an ideal top-hat beam with $D_{en}$ diameter for the entrance pupil, and an ideal gaussian beam with $D_{ex}$ $1/e^2$ diameter for the exit pupil. For the rest of this study, we choose
\begin{equation*}
    D_{en} = D_{ex} = 10\,\text{mm},
\end{equation*}
as arbitrary numbers. 

Solving Eq.\,(\ref{eq:guyon2003}) with $D_{en}$ and $D_{ex}$ gives the results presented in Fig.\,\ref{fig:M1_M2}. The obtained sag curves correspond to an on-axis propagation in the PIAA. See Sec.\,\ref{sec:discussion} for the discussion on the on-axis and off-axis cases.

\section{Results}\label{sec:results}
In this section, we first show the results of the apodization from the PIAA model, and the calculated coupling efficiency with the fibers.

\subsection{Beam apodization with $\partial$Lux}
\begin{figure}
    \centering
    \includegraphics[width=0.5\linewidth]{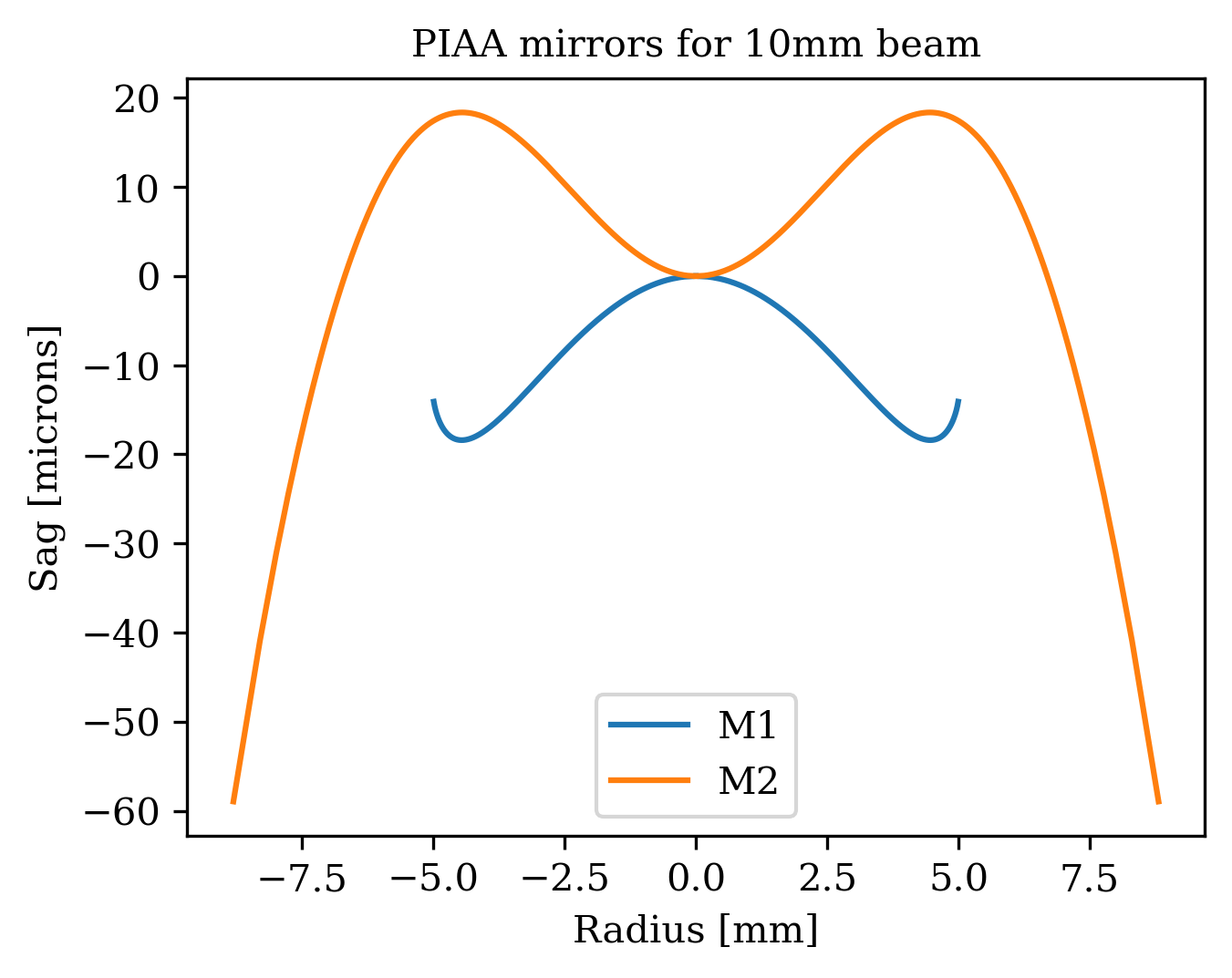}
    \caption{Sag curves of $M_1$ and $M_2$ as obtained from Eq.\,(\ref{eq:guyon2003}) for $D_{en} = D_{ex} = 10\,\text{mm}$. The curves are oriented so that the incoming beam comes from the bottom of the plot for each surface.}
    \label{fig:M1_M2}
\end{figure}

\begin{figure}
    \centering
    \includegraphics[width=0.9\linewidth]{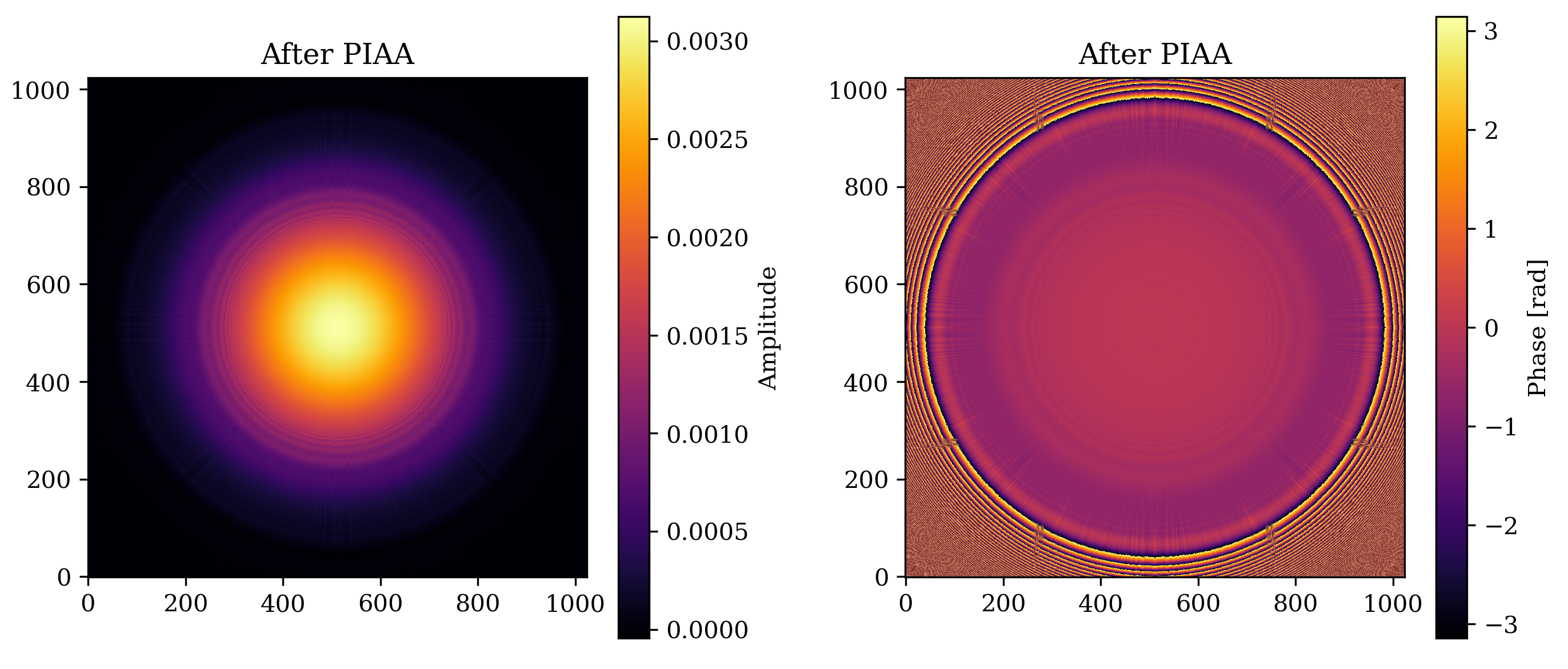}
    \caption{2D wavefront maps of the apodized beam given by the PIAA model, with its normalized amplitude (left), and its phase (right).}
    \label{fig:2D_pupil}
\end{figure}

\begin{figure}
    \centering
    \includegraphics[width=0.9\linewidth]{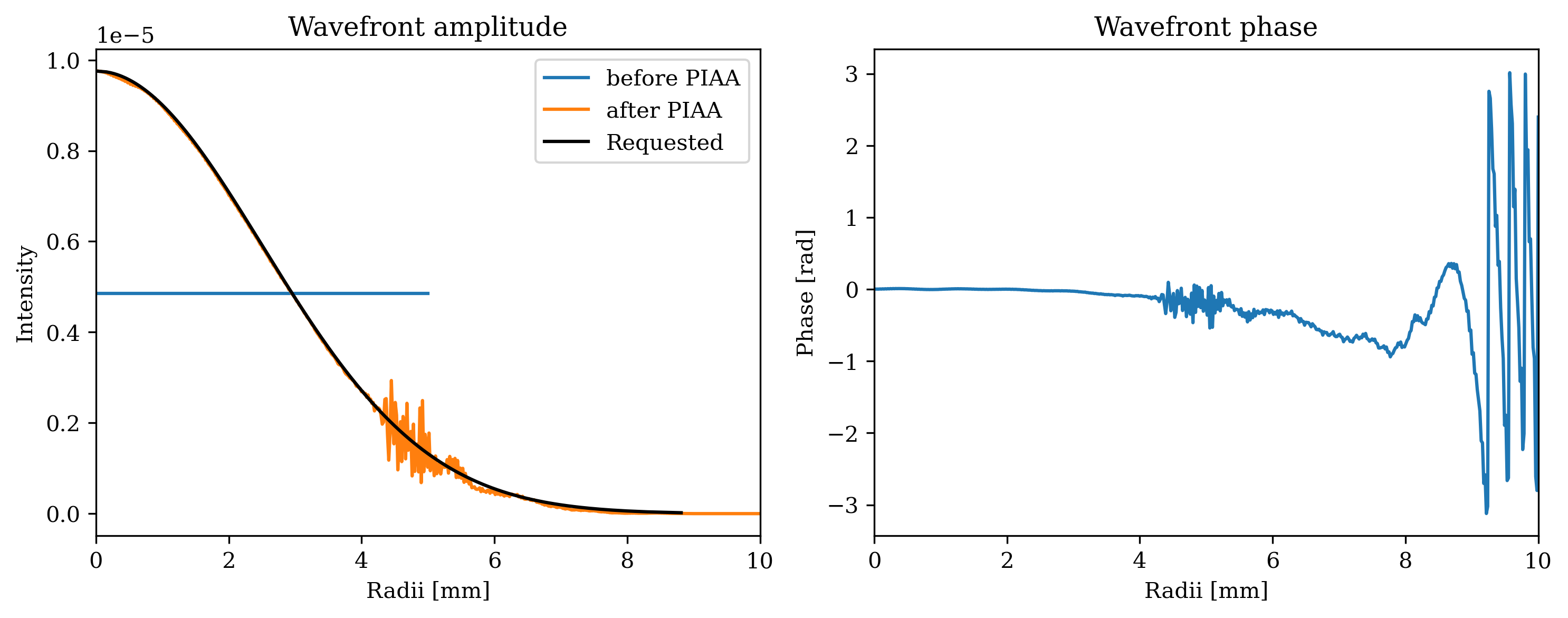}
    \caption{Half cross sections of the wavefront of the apodized beam given by the PIAA model, with its normalized amplitude (left), and its phase (right).}
    \label{fig:1D_pupil}
\end{figure}

We simulate the beam propagating through the PIAA object generated from the $M_1$ and $M_2$ sag curves shown in Fig.\,\ref{fig:M1_M2}. Figures\,\ref{fig:2D_pupil}-\ref{fig:1D_pupil} show the outcome of the PIAA system and the wavefront maps of the apodized beam. Since the PIAA optics are mirrors, the diameter of the apodized beam is achromatic.

\subsection{Coupling efficiency}
\begin{figure}
    \centering
    \includegraphics[width=0.5\linewidth]{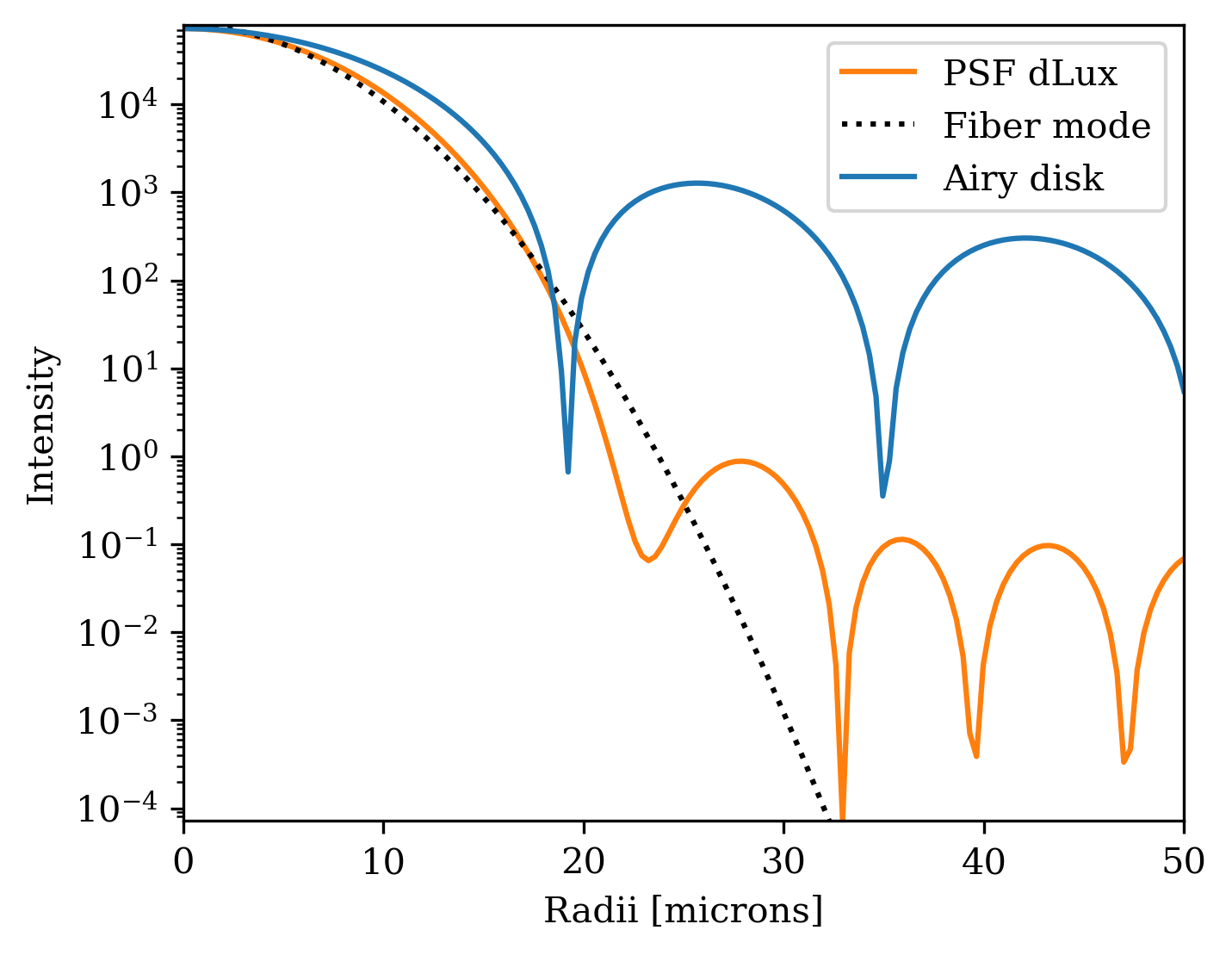}
    \caption{Half cross sections of the fiber mode (black dotted curve), the PSF of the apodized beam from $\partial$Lux (orange curve), and the non-apodized ``Airy disk" beam for comparison (blue curve). The results are obtained at $\lambda=4\,\mu$m, for $\text{MFD}=20\,\mu$m, and $f$ calculated from Eq.\,(\ref{eq:focal_length}).}
    \label{fig:1D_fiber_mode}
\end{figure}

For each wavelength $\lambda$ considered, we use the estimated mode field diameter (MFD) provided for the two fibers TAS \citep{Faber2006} and AgBr \citep{Flatscher2017}. We can then choose the focal length $f$ of the imaging optics to maximize the coupling efficiency of the apodized beam at this wavelength using
\begin{equation}\label{eq:focal_length}
    f = \frac{\pi D_{ex}}{4\lambda}\text{MFD}.
\end{equation}
Figure\,\ref{fig:1D_fiber_mode} gives an example of the point-spread functions (PSFs) and fiber mode in the image plane for $\text{MFD}=20\,\mu$m, and using Eq.\,(\ref{eq:focal_length}) at $\lambda=4\,\mu$m. 


The coupling efficiency $C$ between the electric field of the fiber mode $E_{\text{mode}}$ and the electric field of the apodized PSF $E_{\text{psf}}$ from $\partial$Lux is calculated in the image plane using
\begin{equation}\label{eq:Ceff_equation}
    C = \frac{|\iint E_{\text{mode}} E_{\text{psf}}^*|^2}{\iint |E_{\text{psf}}|^2 \iint |E_{\text{mode}}|^2}.
\end{equation}
The coupling efficiency of an Airy disk in a single-mode fiber should not exceed $\sim82\,\%$ due to geometrical mismatch. Using PIAA, this geometrical limitation can be bypassed and the coupling efficiency can theoretically approach 100\,\%.

\begin{figure}
    \centering
    \includegraphics[width=\linewidth]{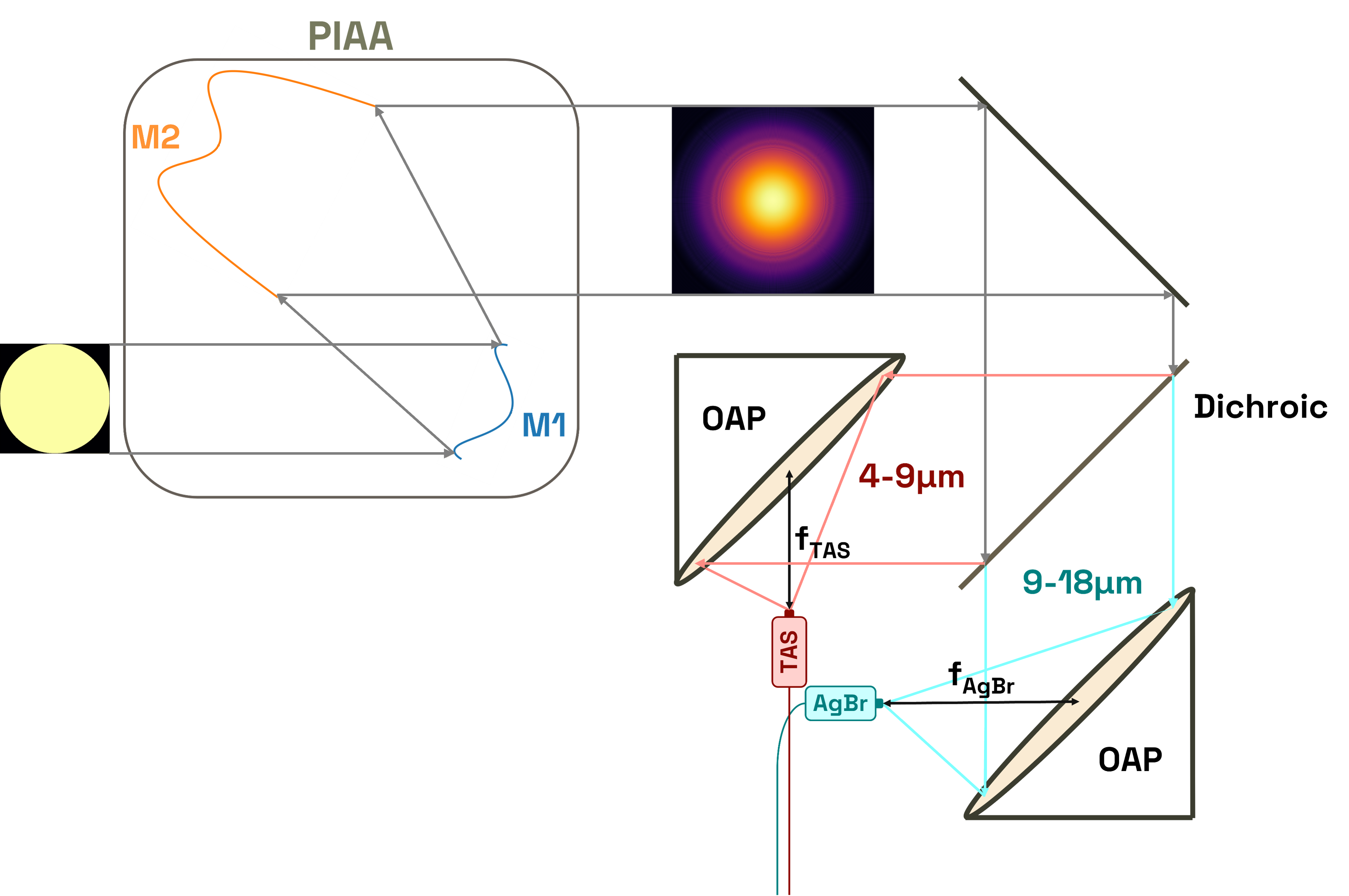}
    \caption{Schematic layout of the PIAA and fiber model. The top-hat beam first propagates through the PIAA optics to be apodized into a gaussian beam. The light is then split in two spectral channels with an ideal dichroic, and injected in the TAS and AgBr fibers using two off-axis parabolas (OAPs) with effective focal lengths of $f_{TAS}$ and $f_{AgBr}$, respectively.}
    \label{fig:PIAA_system}
\end{figure}

Realistically, injecting the light with an off-axis parabola (OAP) for each of the fibers means that its focal length is achromatic, with one value for each fiber. We therefore choose the focal length for the TAS fiber $f_{TAS}$ and the silver halide fiber $f_{AgBr}$ to optimize the broadband coupling efficiency of their associated wavelength range. Figure\,\ref{fig:PIAA_system} gives an overview of the optical system that is modeled.

\begin{figure}
    \centering
    \includegraphics[width=0.5\linewidth]{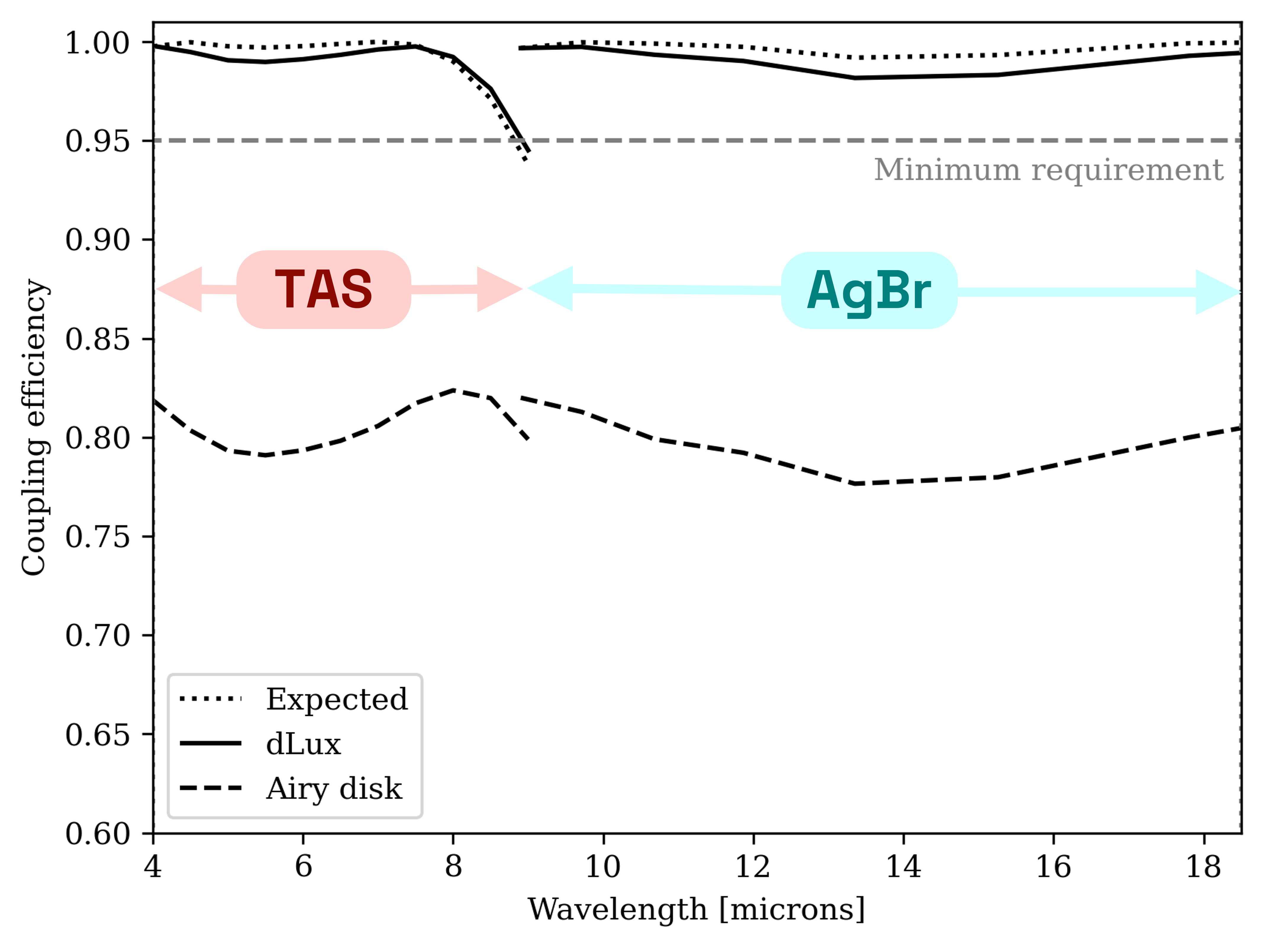}
    \caption{Calculated coupling efficiencies at the entrance of the TAS and AgBr fibers using Eq.\,(\ref{eq:Ceff_equation}). The dashed curve represents the theoretical coupling efficiency without apodization, the dotted curve is the theoretical efficiency for an ideal gaussian beam. The continuous curve is the computed efficiency using the output of the PIAA model from $\partial$Lux. The results are for $f_{TAS}=46\,$mm, and $f_{AgBr}=18\,$mm.}
    \label{fig:Ceff_ideal}
\end{figure}

Figure\,\ref{fig:Ceff_ideal} shows the results of the broadband coupling efficiency for $f_{TAS}=46\,$mm and $f_{AgBr}=18\,$mm. The apodized beam simulated by $\partial$Lux has a coupling efficiency $>95\,\%$ over the full LIFE waveband as required.


\subsection{PIAA optics tolerancing}
The first tolerancing focused on the precision needed for the manufacturing of $M_1$ and $M_2$ to keep the coupling efficiency $>95\,\%$. This is done by adding different kinds of deviations to the nominal mirrors' sag.

\begin{figure}
    \centering
    \includegraphics[width=0.49\linewidth]{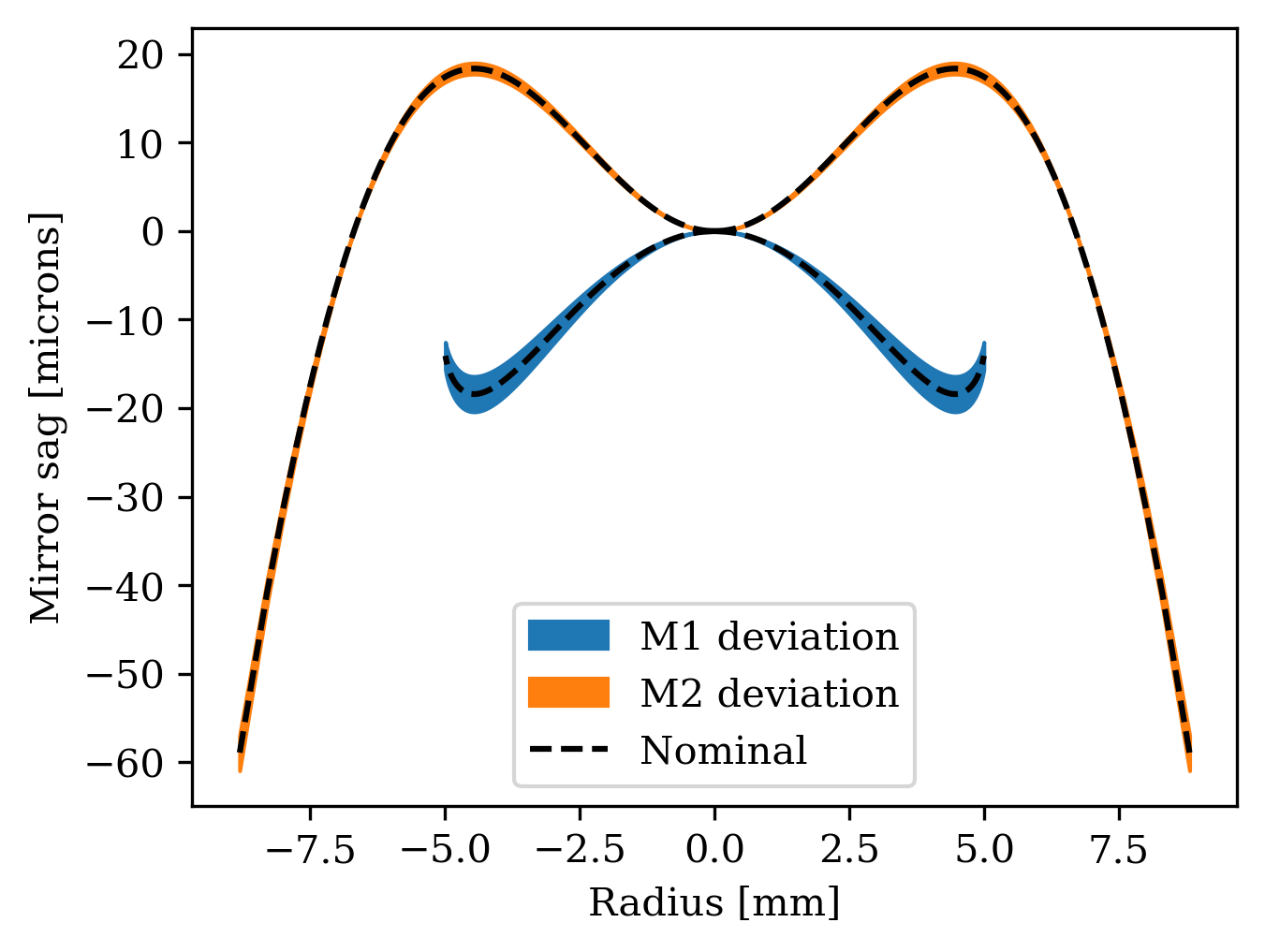}
    \includegraphics[width=0.49\linewidth]{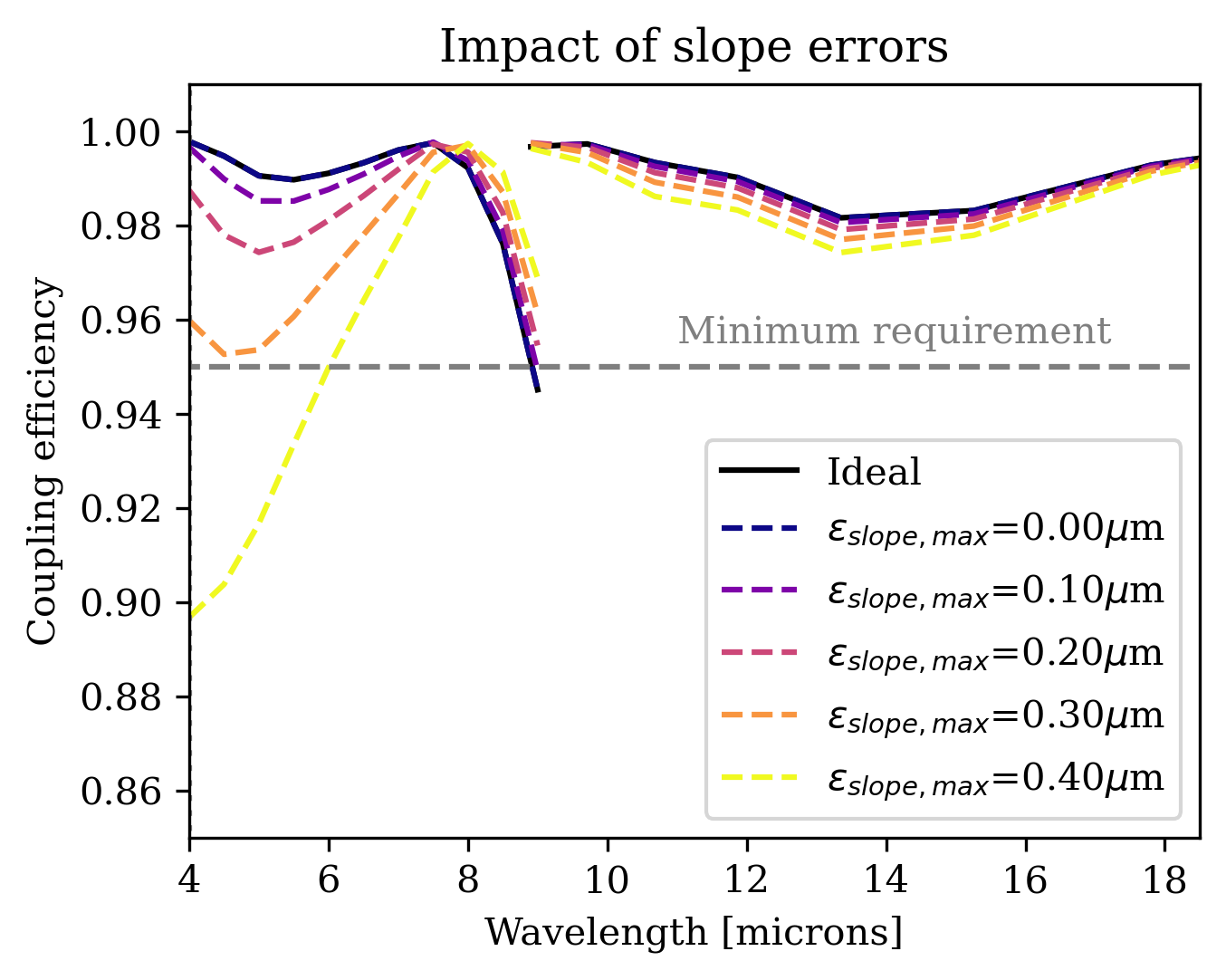}
    \caption{Left: example of slope error for $M_1$ and $M_2$ for $\epsilon_{\text{slope,max}}=2\,\mu$m. Right: calculated coupling efficiencies for perturbed $M_1$ and $M_2$ for different values of $\epsilon_{\text{slope,max}}$.}
    \label{fig:slope_error}
\end{figure}

We call the first deviation ``slope'' error $\epsilon_{\text{slope}}$, where $\epsilon_{\text{slope}}$ is proportional to the nominal sag value $S$
\begin{equation}
    \epsilon_{\text{slope}} = \pm \alpha S,
\end{equation}
where $\alpha$ is chosen so that the maximum value of $\epsilon_{\text{slope}}$ corresponds to the provided $\epsilon_{\text{slope,max}}$.
Figure\,\ref{fig:slope_error} shows an example of slope error on the sag of the PIAA mirrors, and the coupling efficiencies for different values of $\epsilon_{\text{slope,max}}$. An error $\epsilon_{\text{slope,max}}<100\,$nm is keeping the broadband coupling efficiency well above the requirement of 95\,\%.

\begin{figure}
    \centering
    \includegraphics[width=0.49\linewidth]{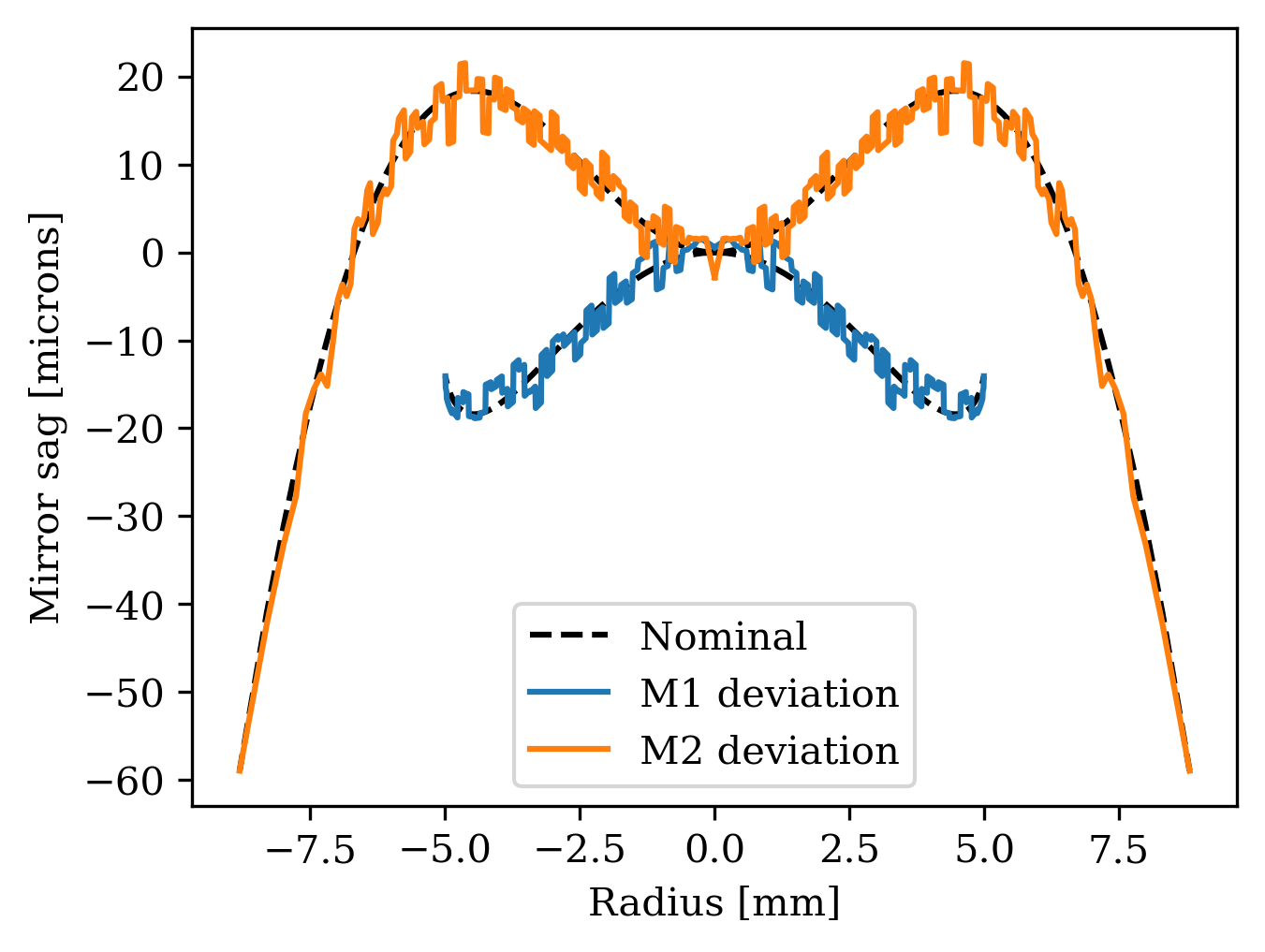}
    \includegraphics[width=0.49\linewidth]{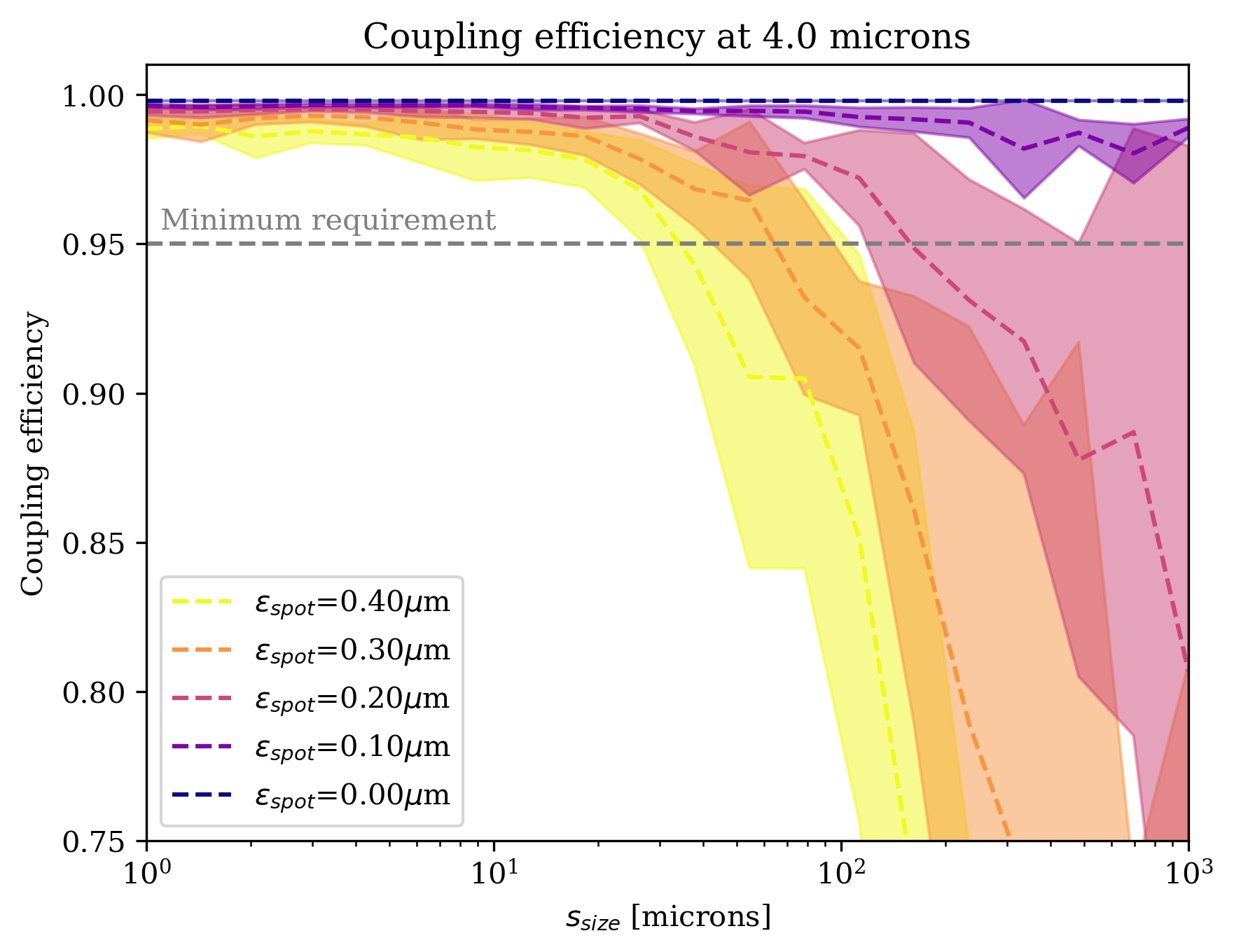}
    \caption{Left: example of spot error for $M_1$ and $M_2$ for $\epsilon_{\text{spot}}=2\,\mu$m and $s_{spot}=0.1\,$mm. Right: calculated coupling efficiencies for perturbed $M_1$ and $M_2$ for different values of $\epsilon_{\text{spot}}$ and $s_{spot}$ at $\lambda=4\,\mu$m.}
    \label{fig:spot_error}
\end{figure}

We call the second deviation ``spot'' error $\epsilon_{\text{spot}}$, where we generate error spots of size $s_{spot}$ with $\epsilon_{\text{slope}}$ as the standard deviation of the spot's error with the nominal surface. 
Figure\,\ref{fig:spot_error} shows an example of spot errors on the sag of the PIAA mirrors, and the coupling efficiency $\lambda=4\,\mu$m for different values of $s_{spot}$ and $\epsilon_{\text{spot}}$. An error $\epsilon_{\text{spot}}<100\,$nm is keeping the coupling efficiency above the requirement of 95\,\% for all values of $s_{spot}$. 

Additional tolerancing is necessary to assess the feasibility of such a solution, especially on the alignment between the beams and the PIAA optics. Previous simulation have shown that off-axis sources quickly suffer from optical aberrations induced by the aspherical mirrors \citep{Guyon2003}.


\section{Discussion}\label{sec:discussion}
In this study, the wavefront propagation and PIAA optics are purely on-axis. To simulate a realistic system, the mirrors either need to be off-axis, or have a central obstruction to let the beam propagate through it. The off-axis surfaces of the PIAA mirrors can be calculated by adding a flat tilt to the on-axis surfaces.

We have assumed the use of step-index single-mode fibers to cover the full LIFE waveband. However, other options are also available to realize spatial filtering in the mid-infrared, especially if the option of silver halide fiber for the longer wavelengths remains uncertain. Other options such as endlessly single-mode photonic crystal waveguides \citep{Ireland2024}, hollow core fibers \citep{Kriesel2011}, pinholes, or photonic chips (Eglin et al. 14154-208 in these proceedings) are also considered as strong options.

We are also only considering the geometric coupling efficiency between the apodized beam and the fiber mode. The efficiency of the fibers will also be impacted by the Fresnel losses at the air-fiber interfaces. They can be attenuated by anti-reflection coatings, but these will likely also reduce the passband of the fiber. The propagation losses will also impact the throughput. The best way to limit these is to improve the efficiency of the coating on the fiber cladding to attenuate the high-order modes, and hence reduce the fiber length.

The location of a PIAA system in LIFE is still under study. One possibility would be to place one on each collector spacecraft to apodize the beam before its propagation to the combiner spacecraft, to reduce the impact of diffraction.

\section{Conclusion}\label{sec:conclusion}
In this work, we simulate the combination of PIAA optics and spatial filtering using step-index single-mode fibers to perform broadband efficient spatial filtering for LIFE. The requirement is to ensure $>95\,\%$ coupling efficiency between 4 and 18.5\,$\mu$m. The solution considered for the fibers is a TAS chalcogenide fiber covering the waveband 4-9\,$\mu$m, and a silver halide fiber covering 9-18.5\,$\mu$m. The simulation shows that the requirement in coupling efficiency can be met using a pair of aspherical mirrors for the PIAA, with manufacturing precision of $<100\,$nm.
This work also shows that the apodization does not need to be chromatic as long as the LIFE waveband is divided into at least two spectral channels covering approximately one octave. 
The next step is now to design and assemble a testbed, and test the performance of PIAA.  

\subsection*{ACKNOWLEDGMENTS} 
This work was supported by the Swiss National Science Foundation [grant number 10004532] and by the Swiss State Secretariat for Education, Research and Innovation (SERI) / Swiss Space Office (SSO).

\bibliographystyle{unsrtnat}
\bibliography{references}  






\end{document}